\documentclass[aps,prl,showpacs,floatfix,superscriptaddress,twocolumn]{revtex4}
\usepackage{graphicx}
\usepackage{amsmath}
\usepackage{amsfonts}
\usepackage{epstopdf}

\begin{document}

\title{Collective motion and nonequilibrium cluster formation in colonies of gliding bacteria}

\author{Fernando Peruani} 
\affiliation{Laboratoire J.A. Dieudonn{\'e}, Universit{\'e} de Nice Sophia-Antipolis, UMR CNRS 6621, Parc Valrose, F-06108 Nice Cedex 02, France}
\affiliation{Max Planck Institute for the Physics of Complex Systems, N\"othnitzer Str. 38, 01187 Dresden, Germany}
\author{J\"orn Starru\ss}
\affiliation{Center for Information Services and High Performance Computing, Technische Universit\"at Dresden, Zellescher Weg 12, D-01069 Dresden, Germany}
\author{Vladimir Jakovljevic}
\affiliation{Max Planck Institute for Terrestrial Microbiology, Karl-von-Frisch Str. 10, D-35043 Marburg, Germany}
\author{Lotte S{\o}gaard-Andersen}
\affiliation{Max Planck Institute for Terrestrial Microbiology, Karl-von-Frisch Str. 10, D-35043 Marburg, Germany}
\author{Andreas Deutsch}
\affiliation{Center for Information Services and High Performance Computing, Technische Universit\"at Dresden, Zellescher Weg 12, D-01069 Dresden, Germany}
\author{Markus B\"ar}
\affiliation{Physikalisch-Technische Bundesanstalt, Abbestr. 2-12, 10587 Berlin, Germany}

\date{\today}

\begin{abstract}
%

We characterize cell motion in experiments and show that the transition to collective motion 
in colonies  of gliding bacterial cells confined to a monolayer appears through 
the organization of cells into larger moving clusters.
Collective motion by nonequilibrium cluster formation is detected for a critical cell packing fraction
 around 17\%. 
This transition is characterized
by a scale-free power-law cluster size distribution, with an exponent $0.88\pm0.07$, and the appearance 
of giant number fluctuations. 
Our findings are in quantitative agreement with simulations of self-propelled rods. 
This suggests that the interplay of self-propulsion 
of bacteria and the rod-shape of bacteria is sufficient to induce collective motion.  
%
%
\end{abstract}

\maketitle

In many microorganisms, the transition from single cell  to multicellular behavior involves
the onset of collective motion, that is characterized by the formation of large cell clusters that move in a coordinated manner.
An open question is by which mechanisms such cellular organization is achieved. 
There are several examples of coordinated cell motion resulting from intercellular signaling systems. 
In the developmental cycle of {\it Dictyostelium discoideum}, a diffusive chemoattractant guides cell motion 
and leads to complex pattern formation~\cite{r10,r20}, while in {\it Myxococcus xanthus} 
a signaling system, that requires cell-to-cell contact, coordinates cell movements and gives rise to rippling patterns~\cite{r30}. 
In absence of  a signaling system, spatial cellular organization can result from 
density dependent diffusivity, as suggested to occur in {\it Escherichia coli} and {\it Salmonella typhimirium}~\cite{tailleur10}.
In other cases,  large-scale coherent patterns are believed to emerge from  hydrodynamic interactions,  as in  swimming bacteria like {\it Bacillus subtilis}~\cite{r35}. 
Thus, collective motion of cellular populations typically involves 
physical and biochemical interactions between cells~\cite{r37}. 
A purely
 physical mechanism has been recently analyzed in simulations of
self-propelled rods~\cite{r32,r300}, in which
the interplay of active rod motion and steric interactions due to
volume exclusion leads to the formation of moving clusters for sufficiently
high densities. Since such collective movement was found to be absent in
the equivalent
equilibrium system of diffusive rods, this phenomenon can be
considered as nonequilibrium cluster formation. Experiments with
granular particles, i.e. artificial self-propelled rods, confirmed
that such a physical mechanism is indeed enough to produce a variety
of collective motion pattern~\cite{r34}. Since many
bacteria are self-propelled and have rod cell shape, simple steric
interactions may be sufficient to induce collective motion even without
the additional impact of biochemical signals.
The aim of this study was to test this idea in experiments using {\it M. xanthus} as model system.

{\it M. xanthus} is a gliding bacterium that has been repeatedly used  to study pattern formation~\cite{r40}, social behavior~\cite{r50}, and motility~\cite{r65}. 
%
%
Locomotion of {\it M. xanthus} involves two different motility systems: the S-motility system, which depends on type IV pili and requires 
cell-to-cell contact~\cite{r70} and the A-motility system that allows individual cells to move~\cite{r80}. 
Force generation by the A-system has been suggested to rely on either slime secretion from the lagging pole~\cite{r112}, or on focal 
adhesion complexes~\cite{r116}. 
Cells occasionally reverse their gliding direction and the reversal frequency is controlled by the {\it frz} chemosensory system~\cite{r90, r100}.  
%


Here, we report on the bacterial self-organization in experiments with a {\it M. xanthus} mutant (SA2407)
which only moves by means of the A-motility system and is virtually unable to reverse.   
%
Complex interactions like social motility mediated by pili and cellular
reversal are  absent in this mutant. 
Hence, the experiments are suitable to test the theoretical hypothesis that collective motion  
can emerge from the  combination of active motion of the cells and 
steric interactions due to volume exclusion~\cite{r32,r300}. 
\begin{figure}
\centering\resizebox{\columnwidth}{!}{\rotatebox{0}{\includegraphics{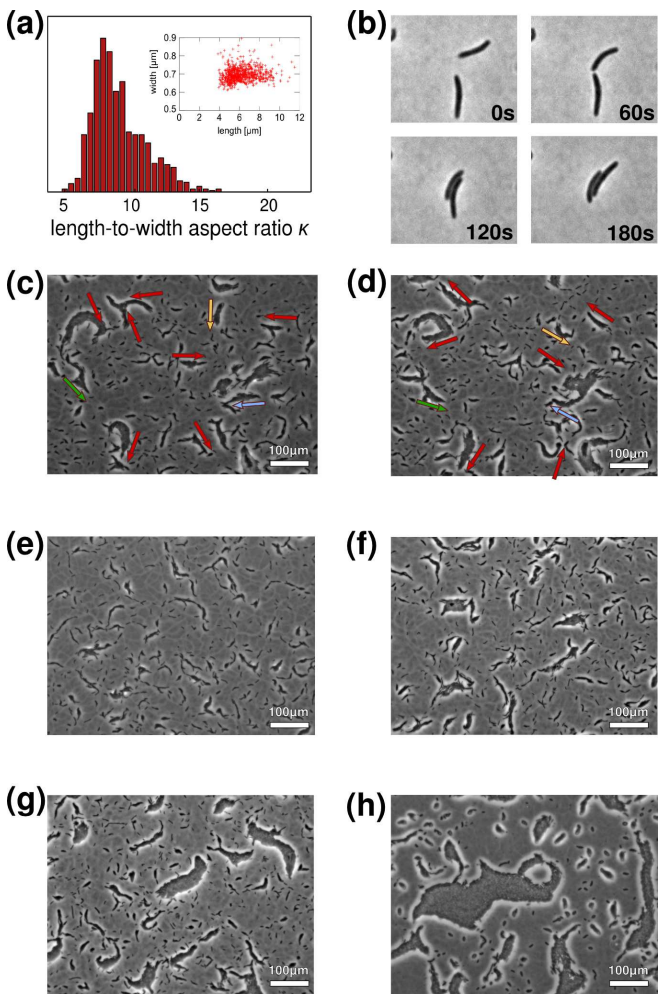}}}
\caption{Clustering of SA2407 cells. (a): Statistics of  cell length-to-width aspect ratio and dispersion of the data in the length-width plane (inset). (b): Cells align their orientations upon collision. (c), (d):  Myxobacterial cells form moving clusters. Arrows indicate the direction of motion of the moving clusters; the time interval between snapshots is 15 min ($\eta =0.11$). (e)-(h): The dynamical clustering process reaches a steady state that strongly depends on cell density. Typical snapshots corresponding  to packing fractions $\eta = 0.06$ in (e) $0.1$ in (f), $0.16$ in (g) and $0.24$ in (h).}\label{fig:snapshots}
\end{figure}

Experiments consisted in spotting a drop of the desired cell density on an agar surface, and subsequently 
the bacterial colony evolution was monitored by taking images of cell arrangements
  every 30 min for a total of 8 hrs. 
To follow the detailed dynamics of cell arrangements, we also made time-lapse recordings of about 30 min, with successive frames taken every 30 seconds. 
Control experiments showed that SA2407 cells do not reverse for the time scale of our experiments, whereas the isogenic {\it frz+} strain reversed with a mean reversal
 period of $\sim 10$ min. 
We found that individual cells glide at an average speed of $v=3.10 \pm 0.35$ $\mu$m/min, and exhibit an average width of about $W=0.7$ $\mu$m and an 
average length of $L=6.3$ $\mu m$, resulting in a mean aspect ratio of $\kappa=L/W=8.9 \pm 1.95$ (Fig.~\ref{fig:snapshots}(a)) and a cell covering an average area $a=4.4$ $\mu m^2$.
Experiments were confined to packing fractions smaller than $0.26$, where the cell dynamics is restricted to a monolayer - 
at larger packing fractions percolation of clusters as well as formation of multilayers are observed. 
The packing fraction $\eta$ is defined as $\eta = \rho\,a$, with $\rho$ the
 (two-dimensional) cell density and $a$ the average covering area of a bacterium given above.

We found that under these conditions over time cells organized into moving clusters. 
Time-lapse recordings showed that collision of cells leads to alignment (Fig.~\ref{fig:snapshots}(b)). 
When the interaction is such that cells end up parallel to each other and move in the same direction, they migrate together for a long time 
(typically $>15$ min)~\cite{comment_antiparallel}. 
Eventually, successive collisions allow a small initial cluster to grow in size. 
In the individual clusters, cells are aligned in parallel to each other and arranged in a head-to-tail manner, as previously described~\cite{r130}. 
In a cluster, cells move in the same direction. 
Cluster-cluster collision typically leads to cluster fusion (Fig.~\ref{fig:snapshots}(c),(d)), whereas splitting and break-up of clusters
 rarely occur. 
On the other hand,  individual cells on the border of a cluster often spontaneously escape from the cluster. 
These two effects, cluster growth due to cluster-cluster collision and cluster shrinkage, mainly due to cells escaping from the cluster 
boundary, compete and give rise to a non-equilibrium cluster size distribution (CSD). 
Typical snapshots of cell arrangements for various packing fraction $\eta$ at the steady state are shown in Figs.~\ref{fig:snapshots}(e)-(h) and reveal
 a strong increase of the cluster size for increasing packing fraction $\eta$.

The CSD - $p(m,t)$ - indicates the probability of a bacterium to be in a cluster of size $m$ at time $t$. 
Note that along the text, the term CSD always refers to this definition. 
The cluster size distribution can be alternatively defined as the number $n_m(t)$ of clusters of size $m$ at time $t$. 
There is a simple relation between these two definitions: $p_m(t) \propto m\,n_m(t)$. 
In experiments we have observed that  the CSD mainly depends on the packing fraction $\eta$. 
Hence, for all snapshots first the packing fraction was determined. 
Then, images with similar packing fraction $\eta$ were compared and the CSD was reconstructed  by determing the CSD for all images within a 
finite interval of the packing fraction. 
The CSD $p(m,t)$ reaches a steady state $p(m)$  after some transient time. 
The duration of this transient depends on the packing fraction $\eta$ and is below 120 min for all $\eta<0.2$, 
see~\cite{si}. 
This indicates that the (non-equilibrium in the thermodynamical sense) clustering process evolves towards a dynamical equilibrium, where the process of  formation of cell clusters of a given size 
is balanced by events in which clusters of this size  disappear by either fusing with other clusters or by loosing individual cells from their boundary.  
%
%
This cluster dynamics is in sharp contrast with cell cluster formation  
 driven by differential cell adhesion and/or cell proliferation as observed, for instance, in cancer cell experiments~\cite{khain2009}, where the CSD never reaches a steady state.

\begin{figure}
\centering\resizebox{\columnwidth}{!}{\rotatebox{0}{\includegraphics{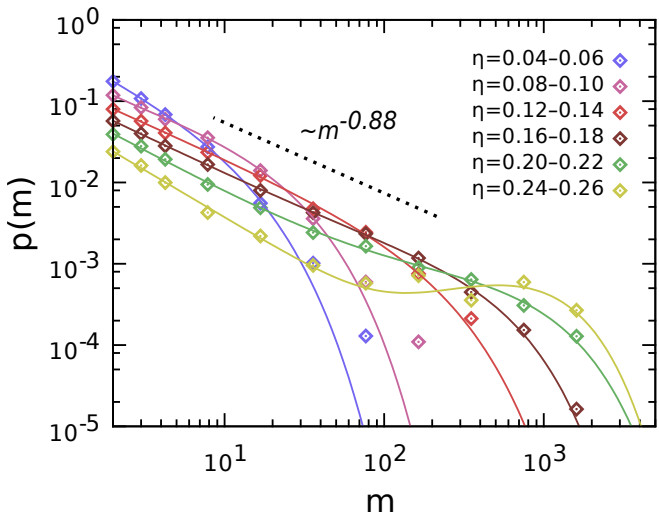}}}
\caption{Cluster statistics of SA2407 cells. The figure shows the steady-state cluster size distribution (CSD) $p(m)$ for various packing fractions $\eta$. Notice the qualitative functional change exhibited by the CSD $p(m)$. 
At the critical point $\eta_c \sim 0.17$, $p(m) \sim m^{\gamma_0}$ with $\gamma_0 = 0.88$. We define this transition point as the onset of collective motion, see text. 
}\label{fig:CSD_experiments}
\end{figure}

The steady-state CSD $p(m)$  strongly depends on the packing fraction $\eta$, with more and more cells moving in larger clusters for increasing packing 
fraction $\eta$.
This is evident in Fig.~\ref{fig:CSD_experiments}, where we observe that at small values of $\eta$, $p(m)$ exhibits a monotonic decay with $m$, 
while at large $\eta$ values, $p(m)$ is non-monotonic, with an additional peak at large cluster sizes. 
%
%
The CSD here was determined at a fixed time (450 minutes) after the beginning of each experiment;  control measurements at other times 
(360 minutes, 540 minutes) revealed practically identical behavior. 
We interpret the presence of a peak at large values of $m$, observed for large $\eta$ values,  as the emergence of collective motion 
resulting in formation of large clusters of bacteria moving in a coordinated fashion. 
This phase is characterized by the existence of large clusters 
that are reflected by the emergence of a local maximum in the CSD, see Fig.~\ref{fig:CSD_experiments}.  
%
%
The transition is evident by the functional change displayed by $p(m)$, monotonically decreasing with $m$ for small values of $\eta$, 
while exhibiting a local maximum at large $\eta$  values. 
At a critical value $\eta_c = 0.17 \pm 0.02$ that separates different regimes of behavior, the CSD can be approximated by
 $p(m) \propto m^{-\gamma_0}$, with $\gamma_0 = 0.88 \pm 0.07$. 
In summary,  for $\eta\leq\eta_c$, the scaling 
of $p(m)$ takes the form:
\begin{equation}\label{eq:scaling}
p(m) \propto  m^{-\gamma_0} \exp(-m/m_0) \, , 
\end{equation}
while for $\eta > \eta_c$, the scaling is:
\begin{equation}\label{eq:scaling_b}
p(m) \propto  m^{-\gamma_1} \exp(-m/m_1) + C m^{\gamma_2} \exp(-m/m_2) \, ,  
\end{equation}
with $\gamma_1,\,\gamma_2,\,m_1,\,m_2$ and $C$ constants that depend on $\eta$. 
In Eq.~(\ref{eq:scaling}), $m_0$ is a function of $\eta$ and increases as $\eta_c$ is approached from below.
Thus, at the critical packing fraction $\eta_c$, $p(m)$ can be approximated by a power-law as long as 
$m$ is much smaller than the total number of cells $N$ in the system.
Eqs.~(\ref{eq:scaling}) and~(\ref{eq:scaling_b}) were obtained in self-propelled rod simulations~\cite{r32, peruani2010} and used here to 
 fit the data in Fig.~\ref{fig:CSD_experiments}.  
%
%
Control experiments with non-motile cells do not exhibit power-law behavior in the CSD for larger packing fraction~\cite{si}. 
Hence, we conclude that without active motion of cells no comparable transition to collective motion occurs. 
In other words, active motion is required for the dynamical self-assembly of cells.
%
%

It is interesting to observe that simulations with self-propelled rods~\cite{r32, peruani2010} exhibit a very similar behavior of $p(m)$.   
%
Moreover,  the exponent $\gamma_0$ takes on similar values as in the experiment. Data in~\cite{r32, peruani2010} give $\gamma_0 = 0.95 \pm 0.05$,   and Yang et al.~\cite{r300} report on similar simulations, obtaining values for $\gamma_0$ in the range from $0.95$ to $1.35$.  
%
%
On the other hand, the mean-field theory for the cluster-size distribution introduced in~\cite{r32} gives an exponent  of $1.3$, which is much larger than the experimental value measured here. 
This theory can be extended to account for elongated rather than circular cluster shapes. 
In the extended theory, predictions for the exponent $\gamma_0$ depend on the functional form of the coagulation and fragmentation kernel. 
Upon a series of assumption, the theory predicts, for elongated clusters, an exponent of $0.85$~\cite{peruani2012}.
%

\begin{figure}
\centering\resizebox{7cm}{!}{\rotatebox{0}{\includegraphics{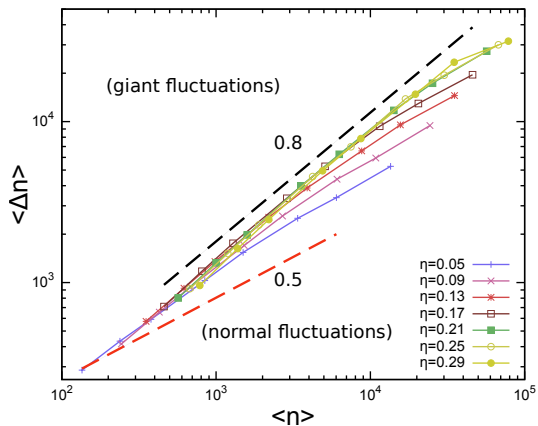}}}
\caption{Number fluctuations of SA2407 cells - at low packing fractions $\eta$, number fluctuations $\Delta n$ are consistent to what is expected for normal fluctuations. For large values of $\eta$, i.e., for $\eta \geq \eta_c$, number fluctuations are giant, with a critical exponent $0.85 \pm 0.05$.}\label{fig:GiantFluctuations}
\end{figure}

We have also characterized the number fluctuations of SA2407 cells:
$\langle \Delta n (l) \rangle = \langle n(l)^2\rangle - \langle n(l)
\rangle^2$,  
where $n(l)$ denotes the number of cells in a box of linear size $l$. 
It can be shown that in general:
%
\begin{equation}\label{eq:GNF}
\langle \Delta n(l) \rangle  \propto \langle n(l) \rangle^{\beta}\, , 
\end{equation}
with $\langle n(l) \rangle = \rho\,l^2$. 
Thus, the quantity $\Delta n$ is a measure of the distribution of cells in space. 
Normal fluctuations correspond to $\beta = 1/2$, while for giant fluctuations $\beta > 1/2$. 
It has been argued that systems of self-propelled particles exhibit in their (orientational) ordered  phase 
 giant number fluctuations, which are often considered a signature of non-equilibrium~\cite{r220}.  
It has been shown recently that self-propelled particles with apolar alignment effectively exhibit such fluctuations~\cite{r240}. 
Fig.~\ref{fig:GiantFluctuations} shows that   $\langle \Delta n \rangle$  is a function of the density $\rho$, respectively, $\eta$. 
At low values of $\eta$, number fluctuations are consistent with normal fluctuations. 
We observe, nevertheless, that for small $\eta$, $\langle \Delta n \rangle$ exhibits a crossover from a regime characterized by 
an exponent close to $0.8$ for small $\langle n \rangle$ to an asymptotic regime for large  $\langle n \rangle$ 
characterized by an exponent $0.5$ as expected for normal number fluctuations. 
As $\eta$ is increased towards $\eta_c$, cells exhibit giant number fluctuations though there is a lack of global orientational order. 
Morevover, for  $\eta \geq \eta_c$ number fluctuations are characterized by the same exponent $\beta=0.8 \pm 0.05$ (Fig.~\ref{fig:GiantFluctuations}).  
Interestingly, this value coincides with the exponent reported in~\cite{r240}. 
These findings suggests that in the experiments  giant number fluctuations are connected  
to  the transition observed in the cluster size statistics.


We have reported on the bacterial cell self-organization  in experiments with a {\it M. xanthus} mutant which only moves by means of the A-motility system and is vitually unable to reverse. 
We found that these bacteria exhibit a transition at a critical packing fraction of $\eta = 0.17$ characterized 
by the emergence of a power-law cluster size distribution, with an exponent $\gamma_0 = 0.88 \pm 0.07$, 
and giant number fluctuations, with exponent $\beta = 0.8 \pm 0.05$, in the absence of global orientational order. 
The observed change in the spatial organization of cells with increasing packing fractions resembles that obtained in simulations with excluded volume interactions  among self-propelled rods~\cite{r32}. 
Moreover, we observe that there is a surprising similarity between the obtained  statistics  
 for the gliding {\it M. xanthus} mutant studied here and the recently 
reported results for the swimming bacterium {\it B. subtilis}~\cite{r1}. 
%
%
These observations raise the questions whether (i) the spatial organization at the onset of collective motion may have universal features and
 (ii) if such features may indeed be linked to the simple physical paradigm of self-propelled rods. 
The latter speculation finds addition tentative support by very recent experimental measurements on the nature and range of
cell-cell interactions in suspensions of {\it E. coli},  where it was found that the emergence of large-scale patterns
in bacterial suspensions is likely to be dominated by simple physical collisions among bacteria  rather than by  
hydrodynamically induced long-range dipole-forces~\cite{r600}. 
Ultimately, only further experimental studies of bacterial colonies and biofilms 
can reveal whether  universal features are  actually present at the onset of collective motion in a wide-range of cellular systems. 
%
%

In summary, we have shown that the cluster size distribution exhibits a qualitative (non-equilibrium) transition (reminiscent to a gelation transition~\cite{ziff1982}) 
from an exponential decaying shape at small $\eta$ to a power-law shape with an additional peak at large $\eta$. 
We have suggested to use this characteristic transition point in the CSD as definition for the onset of collective motion.  
According to this definition, collective motion implies the formation of large moving clusters, with cells 
sharing the same moving direction inside the clusters.
Since in the experiments clusters do not exhibit a moving directional preference, myxobacteria exhibit collective motion 
without global orientational ordering. 
The reported transition to collective motion (via clustering) is hence qualitatively different 
from the transition to global orientational order 
reported in the Vicsek model and its variants~\cite{vicsek1995, peruani2008, r240}. 
It would be interesting to explore the relation between both transitions, i.e., via  clustering and global orientational order, in such minimal models where recently nonequilibrium 
cluster formation similar to the one observed here for colonies of gliding bacteria or for simulations of hard rods has been reported~\cite{peruani2010, huepe2004}. 
%
%
Moreover, we have found that the number fluctuations in the experiments are normal in the limit of 
large numbers $\langle n \rangle$ for densities below the onset of collective motion. 
In contrast, above the transition giant number fluctuations are found. 
The cluster-size statistics as well as the number fluctuation  show both a distinct qualitative change at the onset of collective motion.
Consequently, both measures are suitable to describe the onset of collective motion in large groups of microorganisms.

Finally, the agreement between the cluster statistics obtained in the experiments and in earlier simulations of self-propelled hard rods~\cite{r32, r300}, suggests that the interplay of active motion and volume exclusion is sufficient to explain the collective behavior of the bacteria considered here.  

We are grateful for financial support to DFG through grant DE842/2 and to the Max Planck Society. 
Partial DFG support by GRK 1558 and SFB 910 is acknowledged.



\begin{thebibliography}{99}
\bibitem{comment_antiparallel} Cell collisions often result in antiparallel arrangement of cells, i.e., the alignment symmetry is apolar.
\end{thebibliography}
\end{document}